\documentclass{ws-p10x7}

\begin{document}

\title{A Measurement of $CP$ Violation in $B^0$ Meson Decays with Belle}

\author{Hiroaki Aihara}

\address{Department of Physics, University of Tokyo,\\ 
Tokyo 113-0033, Japan\\
E-mail:aihara@phys.s.u-tokyo.ac.jp\\
{\large \rm Representing the Belle Collaboration}\\
{To appear in the Proceedings of the XXXth International
Conference on High Energy Physics July 27 - August 2, 2000
Osaka, Japan}
}

\twocolumn[\maketitle\abstract{
We present a preliminary measurement of 
the Standard Model $CP$ violation parameter
$\sin 2\phi_1$ using the Belle detector.
A $6.2~{\rm fb}^{-1}$  sample of events
produced by the KEKB asymmetric $e^+e^-$ collider
operating at the $\Upsilon(4S)$ resonance is used.
One neutral $B$ meson is fully reconstructed 
via its decay to a $CP$ eigenstate:
$J/\psi K_S$, $\psi(2S) K_S$, $\chi_{c1} K_S$, $J/\psi K_L$ or $J/\psi \pi^0$.
The flavor of the accompanying $B$ is identified 
mainly from the charges of high momentum leptons or kaons
among its decay products.
The time interval between the two decays is 
determined from the distance between the decay vertices.
A maximum likelihood fitting method is 
used to extract $\sin 2\phi_1$ from
the asymmetry in the time interval distribution.
We report a preliminary result of
$$\sin 2\phi_1=0.45^{+0.43}_{-0.44}({\rm stat})^{+0.07}_{-0.09}({\rm syst}).$$
}]

\section{Introduction}
The Belle experiment at the 
KEKB asymmetric energy $e^+e^-$ $B$-meson factory recently
completed a successful first year of operation.
The KEKB luminosity, which 
was about $10^{31}\ {\rm cm}^{-2}{\rm s}^{-1}$ at the time 
of startup in June 1999, reached a level of $2\times 10^{33}\
{\rm cm}^{-2}{\rm s}^{-1}$ by July 2000.
The Belle detector collected a 
total integrated luminosity of $6.8~{\rm fb}^{-1}$, 
$6.2~{\rm fb}^{-1}$ on the $\Upsilon(4S)$ resonance 
and $0.6~{\rm fb}^{-1}$ off resonance.
We use this data sample to make a preliminary 
measurement of the Standard Model $CP$
violation parameter $\sin 2\phi_1$ using $B_d^0\to J/\psi K_S, \psi(2S)K_S,
\chi_{c1}K_S, J/\psi\pi^0$ and $J/\psi K_L$ decays\footnote{Throughout this paper, when a mode is quoted 
the inclusion of a charge conjugate mode is implied unless otherwise stated.}.

The Standard Model predicts a $CP$ violation through a mechanism of
Spontaneous Symmetry Breaking of the electroweak symmetry 
that results in the Cabibbo-Kobayashi-Maskawa (CKM) quark 
mixing matrix.
In systems of mesons containing a $b$ quark,
$CP$ violating effects are expected to be large.
The interference between the direct $B_d^0\to f_{CP}$ decay amplitude and
the mixing-induced $B_d^0\to \overline{B_d}^0\to f_{CP}$ decay amplitude, 
where $f_{CP}$ is a $CP$ eigenstate
to which both $B_d^0$ and $\overline{B_d}^0$ can decay,  gives rise to an
asymmetry in the time-dependent decay rate:
\setcounter{equation}{0}
\begin{equation}
\label{eq:asymmetry}
\begin{array}{lcl}
A(t)&\equiv&\frac{dN/dt(\overline{B}^0_{t=0}\to f_{CP})-dN/dt({B^0}_{t=0}\to f_{CP})}
{dN/dt(\overline{B}^0_{t=0}\to f_{CP})+dN/dt({B^0}_{t=0}\to f_{CP})}\\
 &=&-\eta_f\sin 2\phi_1 \sin\Delta m_d t,
\end{array}
\end{equation}
where $t$ is the proper time,
$dN/dt(\overline{B}^0_{t=0}$ $(B^0_{t=0})$ 
$\to f_{CP})$ is the decay rate
for a $\overline{B}^0(B^0)$ produced at $t=0$ to decay to $f_{CP}$ at time $t$,
$\eta_f$ is a $CP$-eigenvalue of $f_{CP}$, 
($\eta_f = -1$ for $J/\psi K_S, \psi(2S)K_S$ and
$\chi_{c1}K_S$, and $+1$ for $J/\psi\pi^0$ and $J/\psi K_L$), 
$\Delta m_d$ is the mass difference between two $B^0$ mass eigenstates, and
$\phi_1$ is one of the three internal 
angles\cite{Sanda} of the CKM Unitarity Triangle, defined as
$\phi_1\equiv \pi-\arg(\frac{-V^*_{tb}V_{td}}{-V^*_{cb}V_{cd}}).$

In $\Upsilon(4S)$ decays, $B^0$ and $\overline{B}^0$ mesons
are pair-produced and remain in a coherent
p-state until one of them decays.
The decay of one $B^0$ meson to a final state $f_1$ at time $t_1$
projects the accompanying $B^0$ meson onto an orthogonal state at that time;
this meson then propagates in time and decays to $f_2$ at time $t_2$.
$CP$ violations can be measured if one of 
$B$ mesons decays to a tagging state, $f_{tag}$, i.e.
a final state unique to $B^0$ or $\overline{B}^0$, at time $t_{tag}$ and 
the other decays to an $f_{CP}$ state at time $t_{CP}$.
A time-dependent asymmetry, $A(\Delta t)$, which is obtained 
by replacing time $t$ in (\ref{eq:asymmetry}) with 
the proper time interval $\Delta t\equiv t_{CP}-t_{tag}$,
can then be observed in $\Upsilon(4S)$ decays. 
Because the $B^0\overline{B}^0$ pair is produced nearly at rest in 
the $\Upsilon(4S)$ center of mass system (cms), 
$\Delta t$ can be determined from the distance 
between $f_{CP}$ and $f_{tag}$ decay vertices in the boost ($z$) direction,
$z_{CP}$ and $z_{tag}$, and
$\Delta t\sim \Delta z/\beta\gamma c$, where 
$\beta\gamma$ is a Lorentz boost factor 
of the $\Upsilon(4S)$ restframe  (equal to 0.425 at KEKB). 

This asymmetry is diluted by experimental factors including 
background
to the reconstructed $f_{CP}$ states,
the fraction of events where the flavor of the $B$
meson is incorrectly tagged $(\omega)$, 
and the resolution of the decay
vertex determination $(d_{res})$:
\begin{equation}
\label{eq:dilution}
A_{observed}=\{\frac{1}{1+B/S}(1-2\omega)d_{res} \}A=D A.
\end{equation}
where $B/S$ is the ratio of background\footnote{Background in (\ref{eq:dilution}) 
is assumed to have no asymmetry.}  to signal 
and  $D(<1)$ is the ``dilution factor.''
The statistical error of $\sin 2\phi_1$ is inversely proportional to $D$:
$\delta \sin 2\phi_1 = \frac{1}{\sqrt{S+B}}\frac{1}{D}.$

\section{The Belle Detector}
Belle is a large-solid-angle magnetic spectrometer~\cite{Belle}.
Charged particle tracking is provided by a 
three-layer, double-sided silicon vertex detector (SVD) 
and a small-cell cylindrical drift chamber (CDC) 
consisting of 50 layers of anode wires, 18 of which are 
inclined at small angles.  The tracking system
is situated in a 1.5~T solenoidal field.
The charged particle acceptance is 
$17^\circ<\theta<150^\circ$, where $\theta$ is
the polar angle in the laboratory frame with respect to the beam axis;
the corresponding cms acceptance is
$\sim 92\%$ of the full solid angle.
The impact parameter resolutions are
$\sigma_{r\phi}^2 = (21)^2+(\frac{69}{p\beta\sin^{3/2}\theta})^2\ \mu{\rm m}$
in the plane perpendicular to the beam axis, and 
$\sigma_z^2=(39)^2+(\frac{51}{p\beta\sin^{5/2}\theta})^2\ \mu{\rm m}$ 
along the beam direction, 
where $p$ is the momentum measured in ${\rm GeV}/c$ and 
$\beta$ is the velocity divided by  $c$.
The transverse momentum resolution is 
$(\sigma_{p_t}/p_t)^2=(0.0019p_t)^2+(0.0034)^2$.
Charged hadron identification is provided by $dE/dx$ measurements in the CDC,
aerogel Cherenkov counters (ACC) and 
a barrel of 128 time-of-flight scintillation counters (TOF).
The $dE/dx$ measurements have a resolution for 
hadron tracks of $\sigma(dE/dx)=6.9\%$
and are useful for $\pi/K$ separation for $p<0.8~{\rm GeV}/c$.
The TOF system has a time resolution 
of $95~{\rm ps}~(rms)$ and provides $\pi/K$
separation for $p<1.5~{\rm GeV}/c$. 
The indices of refraction of the ACC elements vary 
with polar angle from 1.01 to 1.03 to match the kinematics
of the asymmetric energy collisions and 
provide $\pi/K$ separation for $1.5~{\rm GeV}/c<p<3.5~{\rm GeV}/c$.
Particle identification probabilities are 
calculated from the combined response of the three
systems.
The efficiency for $K^\pm$ is $\sim 80\%$ with a charged pion 
fake rate of $\sim 10\%$ for all momenta 
up to  $3.5~{\rm GeV}/c$. 

An array of 8736 CsI(Tl) crystals provides 
electromagnetic calorimetry that covers the same 
solid angle as the charged particle tracking system.
The photon energy resolution, estimated from beam tests, is 
$(\sigma_E/E)^2=(0.013)^2+(0.0007/E)^2+(0.008/E^{1/4})^2$,
where $E$ is measured in GeV.
Neutral pions are detected via their decay to $\gamma\gamma.$
The $\pi^0$ mass resolution varies slowly with energy, 
averaging $\sigma_{m_{\pi^0}}=4.9~{\rm MeV}/c^2$.
With a $\pm 3\sigma$ mass selection requirement, 
the overall detection efficiency, including 
geometric acceptance,  for
$\pi^0$s from $B\overline{B}$ events is $40\%$. 
Electron identification is based on a combination of the
CDC $dE/dx$ information,
the response of the ACC, and the position, shape and energy 
deposit of its associated CsI shower.
The electron identification efficiency is above $90\%$ 
for $p>1.0~{\rm GeV}/c$ with a pion fake rate that is below $0.5\%$. 

The magnetic field is returned via an instrumented iron yoke consisting of
alternating layers of resistive plate counters and 4.7~cm thick iron plates.
The total iron thickness of 65.8~cm  plus the 
material of the CsI calorimeter corresponds 
to 4.7 nuclear interaction lengths at normal incidence.
This system, called the KLM, detects muons and $K_L$ mesons in the region 
of $20^\circ<\theta<155^\circ$.
The overall muon identification efficiency is 
above $90\%$ for $p>1~{\rm GeV}/c$
tracks detected in the CDC; a pion fake rate is below $2\%$.
$K_L$ mesons are identified by the presence of the KLM hits
originating from  hadronic interactions of
the $K_L$ in the  CsI and/or iron.
If there are CsI hits associated with a candidate
$K_L$, its  direction is determined from the energy-weighted
center of gravity of the CsI hits alone. Otherwise, the direction
is determined from the average position of the associated KLM hits. 
The angular resolution of the $K_L$ direction 
is estimated to be $\sim 1.5^\circ$
and $\sim 3^\circ$
with and without associated CsI hits.
\begin{figure}[t]
\epsfxsize160pt
\figurebox{120pt}{160pt}{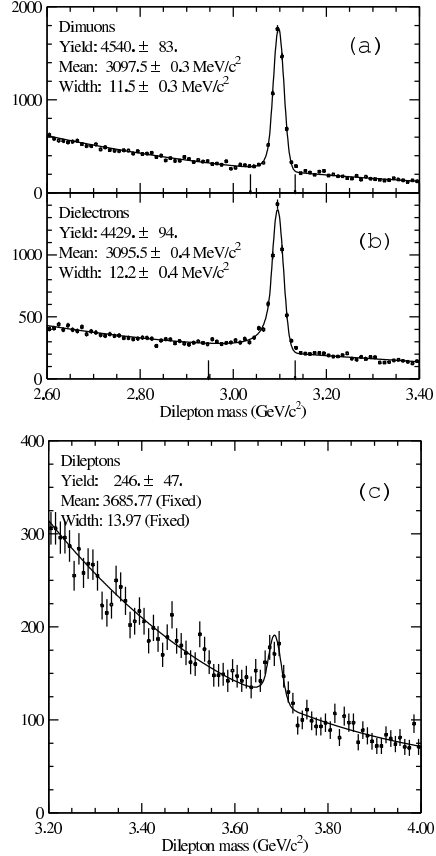}
\caption{The invariant mass distributions for (a) $J/\psi\to \mu^+\mu^-$,
(b) $J/\psi\to e^+e^-$ and (c) $\psi(2S)\to \mu^+\mu^-,e^+e^-$.}
\label{fig:dilepton}
\end{figure}
 
\section{Selection of $B^0$ Decays to $CP$ Eigenstates}
We reconstruct  $B^0$ decays to the following $CP$ eigenstates\cite{Schrenk}:
$J/\psi K_S$, $\psi(2S)K_S$, $\chi_{C1}K_S$ for $CP-1$ states and
$J/\psi \pi^0$,  $J/\psi K_L$ for $CP+1$ states.
The $B^0\overline{B}^0$ hadronic events are selected by requiring 
(i) at least
three tracks with a minimum $p_t$ of $0.1~{\rm GeV/c}$ originating 
within 2.0~cm  and 4.0~cm of the run-by-run average interaction point (IP)
in the plane transverse to the beam axis ($xy$ plane) 
and  along the beam axis ($z$ axis),
respectively;
(ii) at least two neutral energy clusters with a minimum energy of $0.1$~GeV
in the barrel CsI calorimeter;
(iii) a sum of all CsI cluster energies that is
between $10~\%$ and $80~\%$ of the cms energy ($E_{\rm cms}$);
(iv) the total visible (charged and neutral) 
energy greater than $20~\%$ of $E_{\rm cms}$;
(v) $|\sum p^{\rm cms}_z|$ less than $50~\%$ of $E_{\rm cms}/c$, 
where $p^{\rm cms}_z$ is the $z$ component of the momentum calculated
in the cms frame;
(vi) a reconstructed event vertex within 1.5~cm and 3.5~cm of the IP
in the $xy$ plane and
along the $z$ axis, respectively.
In addition, we apply an event topology cut,
$H_2/H_0\leq 0.5$, where $H_2$ and $H_0$ are
the second and zeroth Fox-Wolfram moments, to reject continuum background.

\begin{figure}[t]
\epsfxsize160pt
\figurebox{120pt}{160pt}{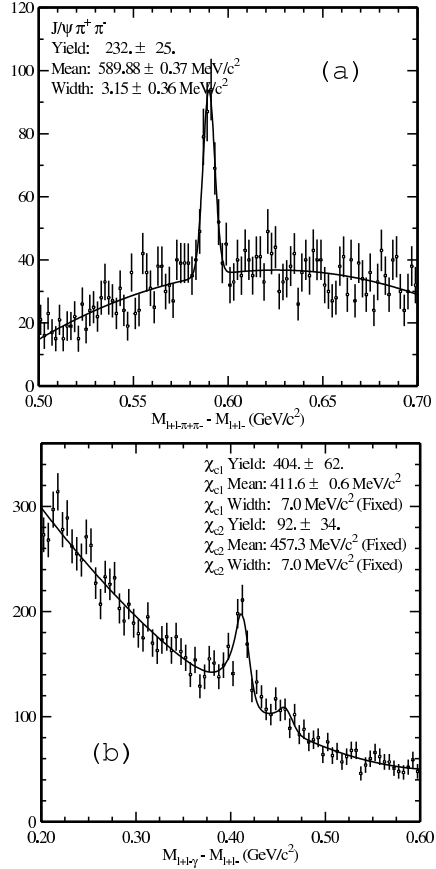}
\caption{the mass difference distributions of (a) $M_{\ell^+\ell^-\pi^+\pi^-}
-M_{\ell^+\ell^-}$ and (b) $M_{\ell^+\ell^-\gamma}-M_{\ell^+\ell^-}$.}
\label{fig:psi2schic1}
\end{figure}

The $J/\psi$ and $\psi(2S)$ mesons are reconstructed via their decays to
$\mu^+\mu^-$ and $e^+e^-$.
Dimuon candidates are oppositely charged track pairs where at least
one track is positively identified as a muon by the KLM system 
and the other is either positively identified as a muon or has 
a CsI energy deposit that is consistent with that of
a minimum ionizing particle.
Similarly, dielectron candidates are oppositely charged 
track pairs  where at least one track is
well identified as an electron and the other track satisfies 
at least the $dE/dx$ or the CsI $E/p$ electron identification requirements.
For dielectron candidates, we correct for
final state radiation or bremsstrahlung in the inner parts of the detector
by including the four-momentum of every photon 
detected within 0.05 radian of the
original electron direction in the $e^+e^-$ invariant mass calculation.
The invariant mass distributions for $J/\psi\to \mu^+\mu^-$,
$J/\psi\to e^+e^-$ and $\psi(2S)\to \mu^+\mu^-,e^+e^-$ are shown in 
Figs.~\ref{fig:dilepton}(a), (b)  and 
(c), respectively.
The $\psi(2S)$ is also reconstructed via its  $J/\psi\pi^+\pi^-$ decay
and the $\chi_{c1}$ is reconstructed through its decay to $J/\psi\gamma$.
Figures~\ref{fig:psi2schic1} (a) and (b) show the mass 
difference distributions  of $M_{\ell^+\ell^-\pi^+\pi^-}
-M_{\ell^+\ell^-}$ and $M_{\ell^+\ell^-\gamma}-M_{\ell^+\ell^-}$.

Candidate $K_S\to \pi^+\pi^-$ decays are oppositely 
charged track pairs that have an invariant mass 
between $482$~and $514~{\rm MeV}/c^2$, 
which corresponds to the $\pm 3\sigma$ around the $K_S$ mass peak.
The $K_S\to \pi^0\pi^0$ decay mode is also used for the
$J/\psi K_S$ channel.
These are selected among photons with a 
minimum energy of 50~MeV and 200~MeV
in the barrel and endcap regions, respectively,
by requiring, assuming $K_S$ decayed at the IP,
(i) a minimum $\pi^0$ momentum of $100~{\rm MeV}/c$;
(ii) $118 <M_{\gamma\gamma}<150~{\rm MeV}/c^2$; and
(iii) $300 <M_{\pi^0\pi^0}<1000~{\rm MeV}/c^2$. 
For each candidate, we determine the most probable $K_S$ decay point by 
minimizing the sum of the $\chi^2$ values from  constraining 
each photon pair to $\pi^0$ invariant
mass while varying the $K_S$ decay point 
along the $K_S$ flight direction defined
by the sum of four photon momenta and the IP.
We then recalculate the invariant masses of the photon pairs and the $\pi^0$s
and require the recalculated $K_S$ mass
to be between $470$~and~$520~{\rm MeV}/c^2$.
For the $J/\psi\pi^0$ mode, the $\pi^0$ candidates are selected 
from photons 
with a minimum energy of 100~MeV.

\begin{figure}[t]
\epsfxsize180pt
\figurebox{120pt}{160pt}{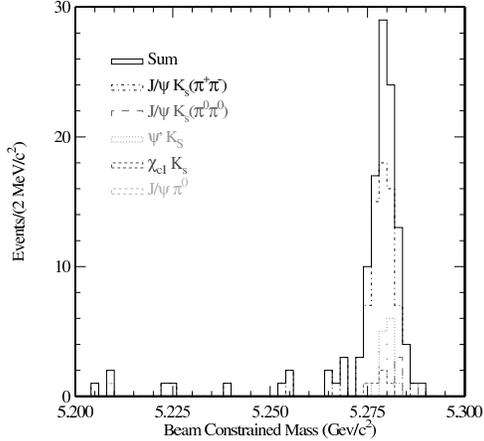}
\caption{The beam-constrained mass distribution of
sum of $B\to$ $J/\psi K_S(\pi^+
\pi^-)$, $J/\psi K_S(\pi^0\pi^0)$, $\psi(2S)K_S(\pi^+\pi^-), \chi_{c1}K_S(\pi^+\pi^-)$,
and $J/\psi \pi^0$ events.}
\label{fig:bmass}
\end{figure}

To identify reconstructed $B$ meson
decays we use the beam-constrained 
mass $M_{beam}\equiv\sqrt{E_{beam}^2-p_B^2}$ and
the energy difference $\Delta E\equiv E_B
-E_{beam}$,
where $E_{beam}$ is $E_{\rm cms}/2$
and $p_B$ and $E_B$ are the $B$ candidate three-momentum and energy calculated 
in the cms.
Figure~\ref{fig:bmass} shows the $M_{beam}$ 
distribution of the combined $B\to$ $J/\psi K_S(\pi^+
\pi^-)$, $J/\psi K_S(\pi^0\pi^0)$, 
$\psi(2S)K_S(\pi^+\pi^-), \chi_{c1}K_S(\pi^+\pi^-)$,
and $J/\psi \pi^0$ samples,
after the imposition of a $~3.5\sigma$ cut 
on $|\Delta E|$ ($\pm 40$~MeV for modes with $K_S\to \pi^+\pi^-$ 
and $\pm 100$~MeV for modes containing $\pi^0$s).
The $B$ meson signal region is defined as 
$|M_{beam}-<M_{beam}>|<0.01~{\rm GeV}/c^2$,
where $<M_{beam}>$ is the mean value of observed $M_{beam}$. 
Table~\ref{tab:tally} lists
the number of signal candidates ($N_{ev}$) and 
the backgrounds ($N_{bkgd}$) determined from
extrapolating the event rate in the
non-signal  $\Delta E$ vs $M_{beam}$ region 
into the signal region,
and from the full Monte Carlo (MC) simulation results. 

\begin{table}[b]
\caption{Summary of the reconstructed $CP$ eigenstates}
\label{tab:tally}
\begin{tabular}{|l|c|c|}
\hline
Mode & $N_{ev}$ & $N_{bkgd}$\\
\hline
$J/\psi K_S(\pi^+\pi^-)$ & 70 & 3.4\\
$J/\psi K_S(\pi^0\pi^0)$  & 4 & 0.3\\
$\psi(2S)K_S(\pi^+\pi^-)$  & 5 & 0.2\\
$\psi(2S)(J/\psi\pi^+\pi^-)K_S(\pi^+\pi^-)$ & 8 & 0.6\\
$\chi_{c1}(\gamma J/\psi) K_S(\pi^+\pi^-)$ & 5 & 0.75\\
$J\psi \pi^0$ & 10 & 1 \\
\hline
Total & 102 & 6.25	\\ 
\hline	 
\end{tabular} 
\end{table}

\begin{figure}
\epsfxsize180pt
\figurebox{120pt}{160pt}{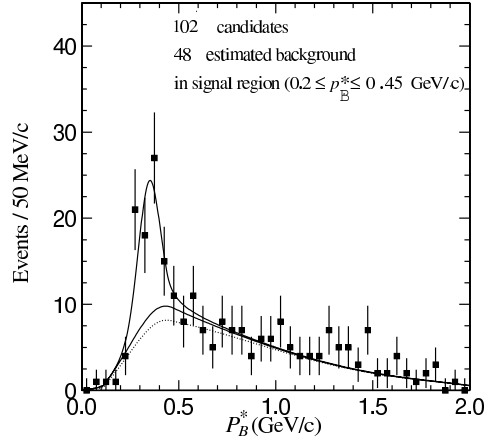}
\caption{The $p_B^*$ distribution with the results of the fit.
The upper solid line is a sum of the signal and background.
The total background (lower solid line) is divided into those coming
from $J/\psi K^{*0}(K_L\pi^0)$ and $J\psi$ + non-resonant $K_L\pi^0$ 
(above the dotted line) and those coming from all other 
sources (below the dotted line). 
}
\label{fig:pbstar}
\end{figure}

The $B^0\to J/\psi K_L$ event candidates are selected by requiring 
the $J/\psi$ momentum and the $K_L$ direction to be consistent with
two-body-decay kinematics.
After requiring the $J/\psi$ cms momentum 
to be between $1.42$~and~$2.0~{\rm GeV}/c$,
we calculate the cms momentum of the $B$ meson, $p_B^*$, 
which should be equal to $\sim 0.34~{\rm GeV}/c$
for a true event.
Figure~\ref{fig:pbstar} shows the $p_B^*$ distribution. 
Also shown are expected distributions of signal
and background derived from the full MC simulation studies.
The background is found to be dominated by $B\to J/\psi X$ events including
$B\to J/\psi K^{*0}(K_L\pi^0)$ and  $B\to J/\psi$ + non-resonant $K_L\pi^0$,
which are a mixture of $CP+1$ and $CP-1$ eigenstates.
There are 102 $J/\psi K_L$ candidates in the signal region 
defined as $0.2\leq p_B^*\leq 0.45~{\rm GeV}/c$.
By fitting the data with the expected shapes, 
we find 48 background events in the signal region, of which 
8 events were from $J/\psi K^{*0}(K_L\pi^0)+J/\psi$ non-resonant $K_L\pi^0$.

\section{Flavor Tagging}
Each event with a $CP$ eigenstate is examined
to see if the rest of the event,
defined as the tag side ($f_{tag}$),
contains a signature specific to $B^0$ or $\overline{B}^0$.
Our tagging methods are based on the correlation 
between the flavor of the decaying $B$ mesons
and the charge of a prompt leptons in $b\to c\nu\ell$ decays,
the charge of Kaon originating from $b\to c\to s$ decays, 
or the charge of $\pi$ from 
$B\to D^*(\to\pi D)\ell\nu$ decays.
A $B^0(\overline {B}^0)$ flavor for $f_{tag}$ 
indicates that $f_{CP}$ was in $\overline {B}^0(B^0)$ state at
$\Delta t=0$.  
We applied the following four tagging methods
in descending order:
if the event failed the method (1), we tested the event with the 
method (2), etc.
\begin{enumerate}
\item High momentum lepton: If $f_{tag}$ contains a 
lepton ($\ell^\pm=e^\pm$ or $\mu^\pm$)  with cms momentum
$p^*\geq 1.1~{\rm GeV}/c$, we assign 
$f_{tag}=B^0(\overline{B}^0)$ for $\ell^+(\ell^-)$.
\item  Charged Kaon: If $f_{tag}$ contains no high 
momentum $\ell^\pm$, the sum of the charges of 
all identified Kaons, $Q_K$, in $f_{tag}$ is determined.
We assign $f_{tag}=B^0(\overline{B}^0)$ if $Q_K>0\ (Q_K<0)$.
If $Q_K=0$, the event fails this method. 
\item Medium momentum lepton: If $f_{tag}$ contains 
an identified lepton in the cms 
momentum range $0.6\leq p_\ell^*<1.1~{\rm GeV}/c$,
we use the cms missing momentum ($p^*_{miss}$) as an
approximation of the $\nu$ cms momentum.
If $p^*_\ell+p^*_{miss}\geq 2.0~{\rm GeV}/c$, we 
assume $f_{tag}$ is from $b\to c\nu\ell$ decay and assign its
flavor based on the charge of $\ell$ as in the method (1).
\item Soft pion: If $f_{tag}$ contains a low 
momentum ($p^*<200~{\rm MeV}/c$) charged track consistent 
with being a $\pi$ from the $D^*\to D\pi$ decay chain,
we assign $f_{tag}=B^0(\overline {B}^0)$ for $\pi^-$ ($\pi^+$).
\end{enumerate}
\begin{figure}
\epsfxsize210pt
\figurebox{120pt}{160pt}{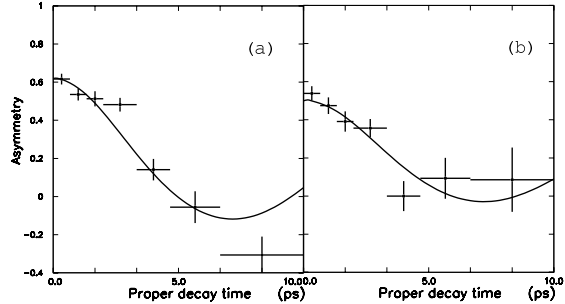}
\caption{The $B^0-\overline{B}^0$ mixing amplitude as a function of
the proper-time interval $\Delta t$ of two neutral $B$ mesons obtained using
(a) $B^0\to D^{*}\ell\nu $ decays and (b) $B^0\to D\ell\nu$ decays.
Also shown are the results of the fit (solid lines).
We obtained $\Delta m_d=0.488\pm 0.026~{\rm pb}^{-1}.$}
\label{fig:wrongtag}
\end{figure}
\begin{table}[t]
\caption{Tagging efficiency ($\epsilon$) and wrong tag fraction $(\omega$)}
\label{tab:tag}
\begin{tabular}{|l|c|c|}
\hline
Method & $\epsilon$ & $\omega$\\
\hline
High $p^*$ $\ell$ & $0.0142\pm 0.021$ & $0.071\pm 0.045$\\
$K^\pm$  & $0.279\pm 0.042$ & $0.199\pm 0.070$\\
Med. $p^*$ $\ell$  & $0.029\pm 0.015$ & $0.29\pm 0.15$\\
Soft $\pi$ & $0.070\pm 0.035$ & $0.34\pm 0.15$\\
\hline	 
\end{tabular} 
\end{table}
The efficiency ($\epsilon$) and the wrong tag fraction ($\omega$)
are determined using a sample of exclusively reconstructed
$B^0\to D^{*-}(D^-)\ell^+\nu$ decays
which are self-tagging decay modes, 
and the full MC simulation.
Because of $B^0-\overline{B}^0$ mixing, the probabilities of finding
the opposite flavor ($OF$) and
same flavor ($SF$) neutral $B$ meson pairs
are
\begin{equation}
\begin{array}{lcl}
P_{OF}(\Delta t)\propto 1+(1-2\omega)\cos(\Delta m_d \Delta t)\\
P_{SF}(\Delta t)\propto 1-(1-2\omega)\cos(\Delta m_d \Delta t).
\end{array}
\end{equation}
Therefore, $\omega$ can be determined from the measurement of 
the $B^0-\overline{B}^0$ oscillation amplitude:
\begin{equation}
A_{mix}\equiv\frac{P_{OF}-P_{SF}}{P_{OF}+P_{SF}}=(1-2\omega)\cos(\Delta m_d\Delta t).
\end{equation}
The flavor of one of the two neutral $B$ mesons is identified
using $B^0\to D^{*-}\ell^+\nu$ where $D^{*-}$ 
decays to $\overline {D}^0\pi^-$ followed by
$\overline{D}^0$ decays to either 
$K^+\pi^-$, $K^+\pi^-\pi^0$, or $K^+\pi^+\pi^-\pi^-$, or
using $B^0\to D^-\ell^+\nu$ where $D^-$ decays to $K^+\pi^-\pi^-$.
We then identify the  flavor of the accompanying $B$ meson
by applying the tagging methods described above.
The vertex position of $D^*\ell\nu$ is determined by requiring
the $\ell$ track and the reconstructed 
$D$ momentum vector form a common vertex.
The vertex position of the tagging $B$ meson is 
determined using the method described in the next section.
The proper-time interval $\Delta t$ of two neutral mesons 
is derived from the distance between the two decay vertices.
We obtain $\omega$ and $\Delta m_d$ from a fit to the $\Delta t$ distributions
of the $OF$ and $SF$ events with the expected functions, which
include the effects of the $\Delta t$ resolution and background.
We determine $\Delta m_d=0.488\pm 0.026~{\rm ps}^{-1}$, 
which is in good agreement with the world average\cite{PDG}.
Figure~\ref{fig:wrongtag} shows the measured 
$A_{mix}$ distributions together with the fitted functions.

Table~\ref{tab:tag} summarizes the tagging efficiency 
and the wrong tag fractions.
The total tagging efficiency is measured to be $0.52$, and
the total effective tagging efficiency ($\epsilon_{eff}$),
 defined as a sum of $\epsilon(1-2\omega)^2$
over all tagging methods, is $0.22$.
The statistical error on $\sin 2\phi_1$ is proportional to 
$1/\sqrt{\epsilon_{eff}}.$ Table~\ref{tab:tag_tally} lists the 
number of tagged events for each $f_{CP}$.
We find a total of 98 tagged events, 
of which 14 events were tagged by a high momentum $e$, 
12 by a high momentum $\mu$,
48 by $K^\pm$, 3 by a medium momentum $e$, 
3 by a medium momentum $\mu$, and 18 by a soft $\pi$. 
\begin{table}
\caption{Summary of tagged events}
\label{tab:tag_tally}
\begin{center}
\begin{tabular}{|l|c|}
\hline
$f_{CP}$ & $N_{ev}$ \\
\hline
$J/\psi K_S(\pi^+\pi^-)$ & 40\\
$J/\psi K_S(\pi^0\pi^0)$  & 4 \\
$\psi(2S)K_S(\pi^+\pi^-)$  & 2\\
$\psi(2S)(J/\psi\pi^+\pi^-)K_S(\pi^+\pi^-)$ & 3\\
$\chi_{c1}(\gamma J/\psi) K_S(\pi^+\pi^-)$ & 3\\
$J\psi \pi^0$ & 4 \\
$J/\psi K_L$ & 42\\
\hline
Total & 98	\\ 
\hline	 
\end{tabular}
\end{center} 
\end{table}
\section{Proper-time Interval Reconstruction}
\begin{figure}
\epsfxsize150pt
\figurebox{120pt}{160pt}{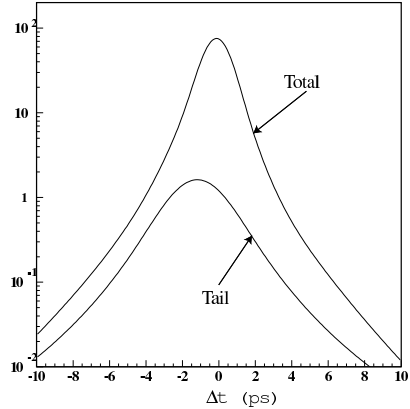}
\caption{The average shape of the event-by-event resolution function
obtained by summing over 300 $B^\pm\to J/\psi K^\pm$ events.}
\label{fig:resolution}
\end{figure}
The $f_{CP}$ vertex  is determined with the two 
lepton tracks from $J/\psi$.
We require that at least one of two tracks must have
SVD hits and that the vertex point is
consistent with the IP profile.
The $f_{tag}$ vertex is formed from the remaining tracks in the event.
In order to reduce the bias due to long-lived particles,
tracks are excluded from the fit: if when combined with 
any other charged track
they form an invariant mass consistent with a $K_S$;
if the track has a large tracking error in $z$ direction
($\sigma_z>0.5~{\rm mm}$);
or if the minimum distance
between the track and the reconstructed $f_{CP}$ vertex is too large,
$\delta z>1.8~{\rm mm}$ or $\delta r>0.5~{\rm mm}$ (in the $r\phi$ plane).
In addition, the track with the worst $\chi^2$ is removed 
from the fit if the reduced $\chi^2$ of the vertex
fit is more than 20.
This procedure is iterated until the reduced $\chi^2$ is below 20.
The  expected vertex resolutions 
are $\sim 40~\mu{\rm m}$ and  $\sim 85~\mu{\rm m}$ 
for the $f_{CP}$ and $f_{tag}$ vertices, respectively.

The most probable $\Delta t$ is estimated as $\Delta z/\beta\gamma c$.
The resolution of $\Delta t$,  $R(\Delta t)$, is parameterized as
a sum of two  Gaussians, a {\it main} Gaussian 
arising from the intrinsic SVD resolution
and the $f_{tag}$ vertex smearing due to the finite 
lifetime of secondary charmed mesons,
and a {\it tail} Gaussian due to a few poorly measured tracks:
\begin{equation}
\begin{array}{lcl}
R(\Delta t)&=&\frac{f_{main}}{\sigma\sqrt{2\pi}}\exp(-\frac{(\Delta  t-\mu)^2}{2\sigma^2})\\
           &+&\frac{f_{tail}}{\sigma_{tail}\sqrt{2\pi}}\exp(-\frac{(\Delta  t-\mu_{tail})^2}{2\sigma_{tail}^2}).
\end{array}
\end{equation}
The mean values ($\mu$, $\mu_{tail}$)  
and widths ($\sigma$, $\sigma_{tail}$) of  the two Gaussians
are calculated event-by-event from the
$f_{CP}$ and $f_{tag}$ vertex errors, taking into account
the error due to the approximation of $\Delta t \sim \Delta z/\beta\gamma c$.
The Gaussian parameters and $f_{tail}(=1-f_{main})$ 
are determined from the full MC simulation studies
and a multi-parameter fit to  $B\to D^*\ell\nu$ data.
In addition, by measuring the lifetime of the  
$D^0\to K^-\pi^+$ decays using only $z$ coordinate
information, we studied the intrinsic $z$ vertex resolution.
In order to show the average shape of the 
$R(\Delta t)$, Fig.~\ref{fig:resolution} was drawn by summing
event-by-event $R(\Delta t)$ functions 
over 300 $B\pm \to J/\psi K^\pm $ (real) events. The width of
$R(\Delta t)$ is dominated by the main Gaussian ($f_{main}=0.96\pm 0.04$);
we find $<\sigma> \sim 1.11~{\rm ps}$, $<\sigma_{tail}>\sim 2.24~{\rm ps}$,
$<\mu>=-0.19~{\rm ps}$, where $<>$ indicates the average over all events.
(The $\mu_{tail}$ was fixed to $-1.25~{\rm ps}$ based on the simulation
studies.) The non-zero negative mean values of the Gaussians reflect 
the bias in $f_{tag}$ vertex position due to secondary charmed mesons.

\begin{figure}
\epsfxsize190pt
\figurebox{120pt}{160pt}{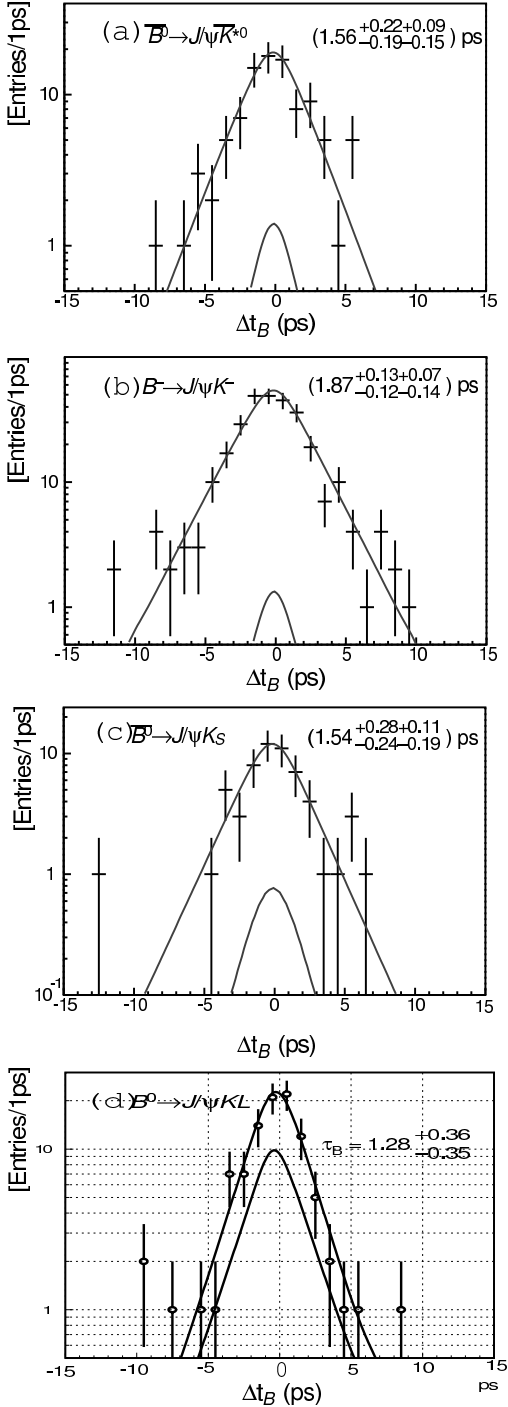}
\caption{$\Delta t$ distributions and the results of the lifetime fit for
(a) $\overline{B}^0\to J/\psi \overline{K}^{*0}$, (b) $ B^-\to J/\psi K^-$, 
(c) $\overline{B}^0\to J/\psi K_S$ and (d) $\overline{B}^0\to J/\psi K_L$.
The lower solid curve represents the background distribution.}
\label{fig:Blife}
\end{figure}

Based on the above described vertex reconstruction and 
resolution function,
we measured lifetimes of neutral and charged $B$ mesons\cite{Tajima}.
Figure~\ref{fig:Blife} shows $\Delta t$ distributions of some of measured 
decay modes
with the results of the lifetime fit.
Table~\ref{tab:Blife} summarizes the results.
The obtained values for different decay modes are consistent with each other 
and are in good agreement with the 
world averages\cite{PDG}, $\tau_{B^0}=1.548\pm 0.032~{\rm ps}$, and
$\tau_{B^\pm}=1.653\pm 0.028~{\rm ps}$.
This verifies the validity of our $\Delta z$ measurement and $R(\Delta t)$.

\section{Extraction of $\sin 2\phi_1$}
An unbinned maximum likelihood method is used to extract the best value 
for $\sin 2\phi_1$.
The probability density function expected 
for the signal distribution with a $CP$ eigenvalue of
$\eta_f$ is given by:
\begin{equation}
\begin{array}{l}
Sig(\Delta t,\eta_f,q)=\frac{1}{\tau_{B^0}}\exp(-|\Delta t|/\tau_{B^0})\\
\times \{1-q(1-2\omega)\eta_f\sin 2\phi_1\sin (\Delta m_d\Delta t)\} ,
\end{array}
\end{equation}
where $q=1(-1)$ if $f_{tag}=B^0(\overline{B}^0)$ and 
$\omega$ depends on the method of the flavor 
tagging as given in Table~\ref{tab:tag}.
\begin{table}
\caption{Summary of $B$ meson lifetime measurements}
\label{tab:Blife}
\begin{tabular}{|l|l|}
\hline
Decay mode & lifetime (ps) \\
\hline
$\overline{B}^0\to D^{*+}\ell^-\nu$ & $1.50\pm 0.06^{+0.06}_{-0.04}$\\
$\overline{B}^0\to D^{*+}\pi^-$ & $1.55^{+0.18+0.10}_{-0.17-0.07}$\\
$\overline{B}^0\to D^+\pi^-$ & $1.41^{+0.13}_{-0.12}\pm 0.07$\\
$\overline{B}^0\to J/\psi \overline{K}^{*0}$& $1.56^{+0.22+0.09}_{-0.19-0.15}$\\
\hline
Combined & $1.50\pm 0.05\pm 0.07$\\
\hline	
$\overline{B}^0\to J/\psi K_S$ & $1.54^{+0.28+0.11}_{-0.24-0.19}$\\
$\overline{B}^0\to J/\psi K_L$ & $1.28^{+0.36}_{-0.35}$\\
\hline
\hline
$B^-\to D^{*0}\ell^-\overline{\nu}$ & $1.54\pm 0.10^{+0.14}_{-0.07}$\\
$B^-\to D^0\pi^-$ & $1.73\pm 0.10\pm 0.09$\\
$B^-\to J/\psi K^-$& $1.87^{+0.13+0.07}_{-0.12-0.14}$\\
\hline
Combined & $1.70\pm 0.06^{+0.11}_{-0.10}$\\
\hline
\end{tabular} 
\end{table} 
The values of $\tau_{B^0}$ and $\Delta m_d$ are 
fixed to the world averages\cite{PDG},
$1.548\pm 0.032~{\rm ps}$ and $0.472\pm 0.017~{\rm ps}^{-1}$, respectively.
By investigating events in background-dominated regions 
(the side bands in the $\Delta E$ vs $M_{beam}$ scatterplot), 
we find that the probability density function for background events
for all $f_{CP}$ events (except for $B^0\to J/\psi K_L$ events)
is consistent with 
$Bkg(\Delta t)=\frac{1}{2\tau_{bkg}}\exp(-|\Delta t|/\tau_{bkg})$, where
$\tau_{bkg}=0.73\pm 0.12~{\rm ps}$. 
The likelihood of an event, $i$, is calculated as:
\begin{equation}
\begin{array}{l}
\rho_i=p_{sig}\int^{+\infty}_{-\infty} Sig(s,\eta_f,q)R(\Delta t - s)ds \\
+(1-p_{sig})\int^{+\infty}_{-\infty} Bkg(s)R(\Delta t -s)ds, 
\end{array}
\end{equation}
where $p_{sig}$ is the probability for 
the event being a signal, and $R(\Delta t)$ is the resolution 
function described in the previous section.
The log-likelihood ${\displaystyle -\sum_i \ln\rho_i}$ is calculated
by summing over all signal events.
The most probable $\sin 2 \phi_1$ value is found
by scanning over $\sin 2 \phi_1$ values
to minimize the log-likelihood function.

\begin{figure}
\epsfxsize200pt
\figurebox{120pt}{160pt}{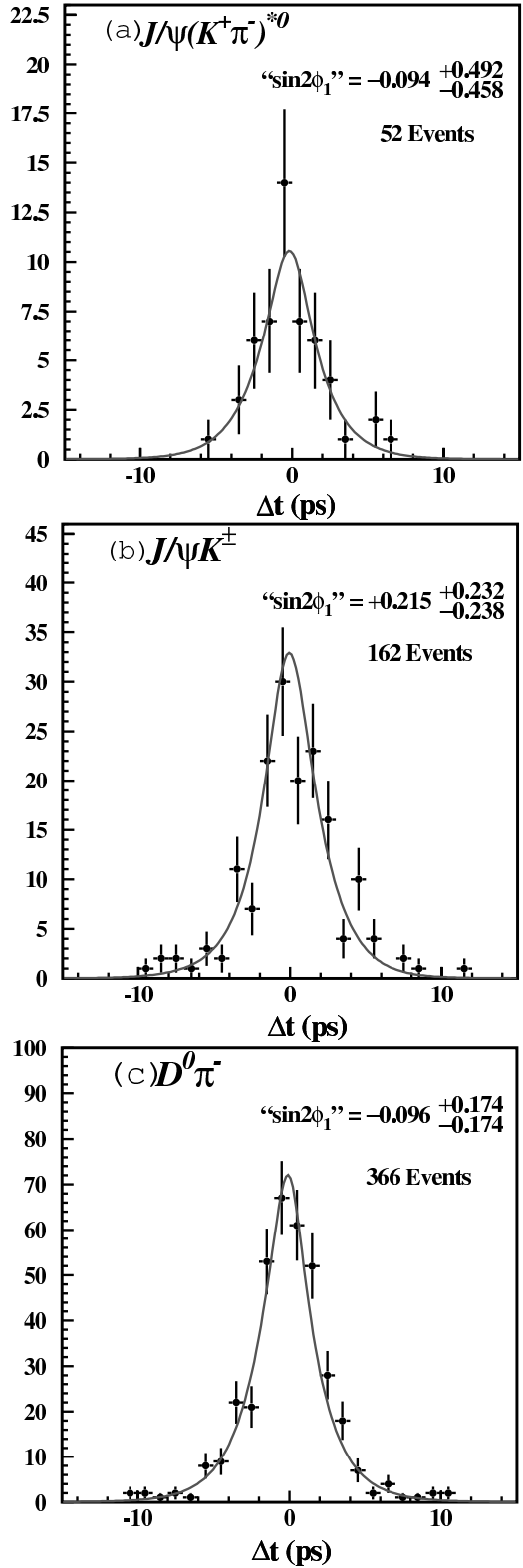}
\caption{$\Delta t$ distributions and the results of the $CP$ fit for
(a) $B^0\to J/\psi K^{*0}(K^{*0}\to K^+\pi^-)$,
(b) $B^- \to J/\psi K^-$, and (c) $B^-\to D^0\pi^-$ events.}
\label{fig:Control}
\end{figure}

To test for possible bias in the analysis,
we apply the same analysis program including 
tagging, vertexing and log-likelihood fitting to control data samples 
with null intrinsic asymmetry: $B^0\to J/\psi K^{*0}(K^{*0}\to K^+\pi^-)$;
$B^- \to J/\psi K^-$; $B^-\to D^0\pi^-$; and $B^0\to D^{*-}\ell^+\nu$ decays.
Figure~\ref{fig:Control} shows the $\Delta t$ 
distributions of $B^0\to J/\psi K^{*0}(K^{*0}\to K^+\pi^-)$,
$B^- \to J/\psi K^-$, and  $B^-\to D^0\pi^-$ events.
The results of the fit for apparent $CP$ asymmetry are 
given in Table~\ref{tab:control}; they
are all consistent with null asymmetry.
\begin{table}
\caption{Results of $CP$ fit to control data.}
\label{tab:control}
\begin{tabular}{|l|l|}
\hline
Decay mode & Apparent $\sin 2\phi_1$ \\
\hline
$\overline{B}^0\to J/\psi (K^+\pi^-)^{*0}$& $-0.094^{+0.492}_{-0.458}$\\
$B^-\to J/\psi K^-$& $+0.215^{+0.232}_{-0.238}$\\
$B^-\to D^0\pi^-$ & $-0.096\pm 0.174$\\
$B^0\to D^{*-}\ell^+\nu$ & $+0.09\pm 0.18$\\
\hline
\end{tabular} 
\end{table}
\begin{figure}
\epsfxsize180pt
\figurebox{140pt}{180pt}{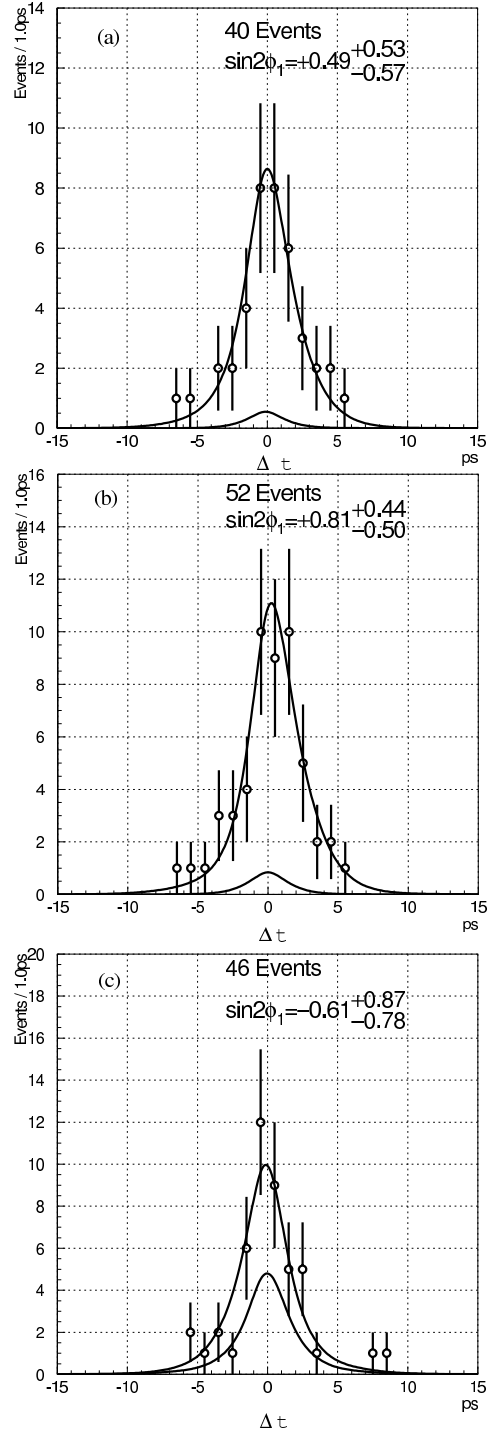}
\caption{$dN/d\Delta t|_{q=+1}+dN/d(-\Delta t)|_{q=-1}$ distributions
and the results of the $CP$ fit for
(a) $B^0\to J/\psi K_S(\pi^+\pi^-)$ only,
(b) all $CP-1$ $f_{CP}$ modes combined, and
(c) all $CP+1$ $f_{CP}$ modes combined.
A lower solid line in each figure indicates background contribution.}
\label{fig:cp}
\end{figure}
\begin{figure}
\epsfxsize180pt
\figurebox{120pt}{160pt}{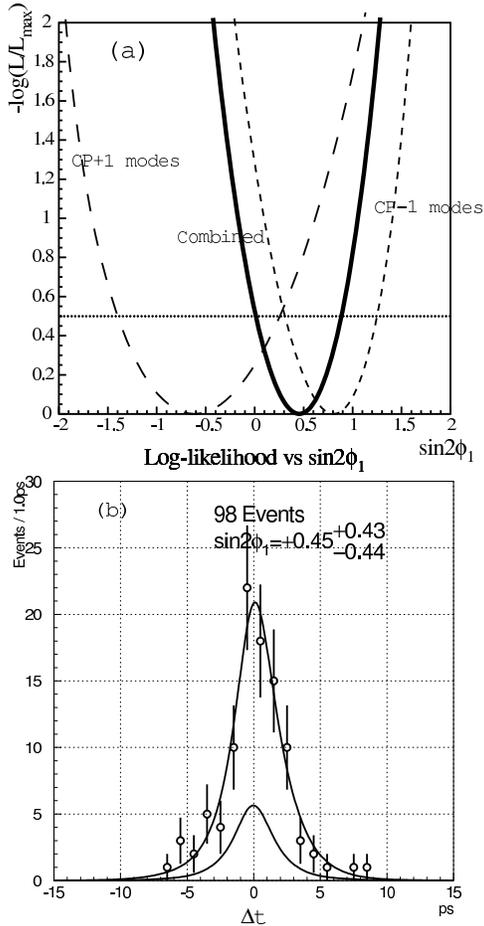}
\caption{The log-likelihood as a function of $\sin 2\phi_1$ (a) and
a sum of $dN/d(+\Delta t)|_{q=+1}$
$+dN/d(-\Delta t)|_{q=-1}$ for 
$\eta_f=-1$ and $dN/d(-\Delta t)|_{q=+1}$
$+dN/d(+\Delta t)|_{q=-1}$ for $\eta_f=+1$ (b) 
for combined $CP-1$ and $CP+1$ events.
}
\label{fig:combined}
\end{figure}
\begin{table}
\caption{Results of $CP$ fit to tagged $f_{CP}$ events.}
\label{tab:CPfit}
\begin{tabular}{|l|l|}
\hline
Decay mode &  $\sin 2\phi_1$ \\
\hline
$J/\psi K_S(\pi^+\pi^-)$ only & $+0.49^{+0.53}_{-0.57}$\\ 
All $CP-1$ modes & $+0.81^{+0.44}_{-0.50}$\\
All $CP+1$ modes & $-0.61^{+0.87}_{-0.78}$\\
\hline
All combined & $+0.45^{+0.43}_{-0.44}$\\
\hline
\end{tabular}
\end{table} 
The results of the fit to the 
tagged $f_{CP}$ events are summarized in Table~\ref{tab:CPfit}.
To display the fitted results, 
the $dN/d\Delta t$ distribution for $q=+1$ events 
and $dN/d(-\Delta t)$ distribution for $q=-1$ events are added:
\begin{equation}
\label{dndt}
\begin{array}{l}
dN/d\Delta t|_{q=+1}+dN/d(-\Delta t)|_{q=-1}\\
\propto \exp(-|\Delta t|/\tau_{B^0})\\
\times \{1-(1-2\omega)\eta_f\sin 2\phi_1\sin (\Delta m_d\Delta t)\}.
\end{array}
\end{equation}
Figures~\ref{fig:cp} (a) and (b) show the results for only 
$f_{CP}=J/\psi K_S(\pi^+\pi^-)$ events 
and  for all  $CP-1$ events combined.  
Figure~\ref{fig:cp} (c) shows the result for $CP+1$ events, 
i.e. $J/\psi K_L$ and $J/\psi \pi^0$.
In fitting to $f_{CP}=J/\psi K_L$ events,  
the background due to $J/\psi K^*(K_L\pi^0)$ $+$
non-resonant  $J/\psi K_L\pi^0$,
which amounts to $\sim 17\%$ of the total background, 
is taken to be a mixture of $CP-1$ (73\%)
and $CP+1$ (27\%) states, based on the results of a
$B\to J/\psi K_S\pi^0$ analysis\cite{paper285}.
A fit to 52 events in the $J\psi K_L$ 
sideband (i.e.  $1.0<p_B^*<2.0~{\rm GeV}/c$ region), 
where the non-$CP$ $J/\psi X$ events  dominate, 
gives the result 
$\sin 2\phi_1=+0.02^{+0.48}_{-0.49}$, consistent with null asymmetry. 
Finally we perform 
the simultaneous fit to $CP-1$ and $CP+1$ events to extract the best 
$\sin 2\phi_1$ value. 
Figure~\ref{fig:combined} (a) shows the log-likelihood  as a function of
$\sin 2\phi_1$ for a total of 98 $CP-1$ 
and $CP+1$ combined events together with
the  results for $CP-1$ and $CP+1$ separately.
We find $\sin 2\phi_1=+0.45^{+0.43}_{-0.44}$.
To display the results of the fit,  
$dN/d(+\Delta t)|_{q=+1}+dN/d(-\Delta t)|_{q=-1}$ for 
$\eta_f=-1$ and 
$dN/d(-\Delta t)|_{q=+1}+dN/d(+\Delta t)|_{q=-1}$ for $\eta_f=+1$ 
are added so that 
the distribution becomes approximately proportional to 
$\exp(-|\Delta t|/\tau_{B^0})\{
1+(1-2\omega)\sin 2\phi_1\sin(\Delta m_d\Delta t)\}$, as shown in
Fig.~\ref{fig:combined} (b).

\begin{figure}
\epsfxsize200pt
\figurebox{120pt}{160pt}{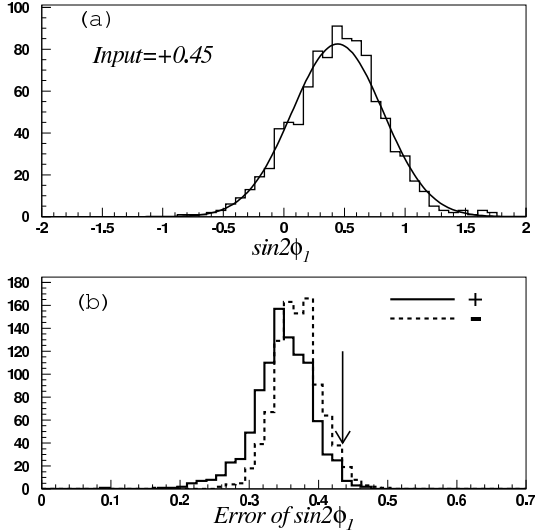}
\caption{Results of 1000 toy Monte Carlo experiments 
generated for $\sin 2\phi_1=0.45$:
distributions of the fitted $\sin 2\phi_1$ (a) and
statistical errors (b). In (b) distributions for errors 
on $+$ side and $-$ side are shown separately.  
Each experiment contains 98 events with the same tag 
composition as the real data.}
\label{fig:toy}
\end{figure}

We generated 1000 toy Monte Carlo experiments with the same 
number of tagged $CP$ events,
having the same composition of the tags 
and the same resolutions as in the $CP$ data sample,
for an input value of $\sin 2\phi_1=0.45$. 
Figures~\ref{fig:toy} (a) and (b)  show the distributions of the central
$\sin 2\phi_1$ value and 
the statistical errors, $+$ side and $-$ side separately.
We found that the probability of obtaining a value of the 
statistical error greater than the observed value was $\sim 5\%$.

Table~\ref{tab:systematics} lists systematic errors.
The largest error is due to uncertainty in the wrong tag fraction
determination.   This was studied by varying 
$\omega$ individually for each tagging method.
The effect due to uncertainty in $\Delta t$ resolutions for both signal and
background was studied by varying parameters in $R(\Delta t)$.
Also included are the effects due to uncertainties in estimate of 
the background  fraction, in the world average $\tau_B$ and $m_d$ values.
An imperfect knowledge of the event-by-event IP 
profile could cause a systematic error in $\sin 2\phi_1$ 
due to the vertex reconstruction.
The effect was studied by repeating the entire fitting procedure  
by varying the IP envelope by $\pm 1 \sigma$ in all 3 dimensions. 
The total systematic error of $\sin 2\phi_1$ is found to be
$+0.07-0.09$. 
\section{Conclusion}
\begin{figure}
\epsfxsize200pt
\figurebox{120pt}{160pt}{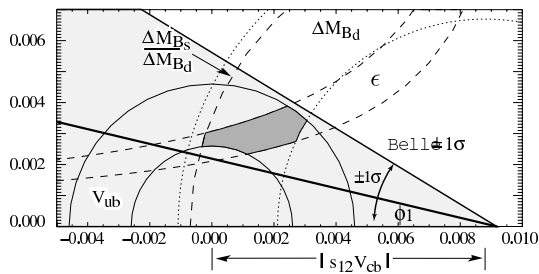}
\caption{The region in the $\rho-\eta$ plane corresponding to this measurement
$\pm 1\sigma$,
$\sin 2\phi_1=0.45^{+0.44}_{-0.45}$.
This was made by modifying Fig. 11.2 of ref. [5].}
\label{fig:rho-eta}
\end{figure}

Based on a $6.2~{\rm fb}^{-1}$ data sample collected 
at the $\Upsilon(4S)$ resonance during the
first year of the KEKB operation, 
we made a preliminary determination of 
$\sin 2\phi_1$ using 98 flavor-tagged
events consisting of 40 $J/\psi K_S(\pi^+\pi^-)$, 4 $J/\psi K_S(\pi^0\pi^0)$,
5 $\psi(2S)K_S(\pi^+\pi^-)$, 3 $\chi_{C1}K_S(\pi^+\pi^-)$, 4 $J/\psi\pi^0$ and
42 $J/\psi K_L$ events. 
We found 
$$\sin 2\phi_1 = 0.45^{+0.43}_{-0.44}({\rm stat})^{+0.07}_{-0.09}({\rm syst}).$$
Figure~\ref{fig:rho-eta} shows the region in the $\rho-\eta$ plane\footnote{The $\rho-\eta$ plane is
the complex plane of the triangle formed by the CKM matrix elements
$V_{ub}^*$, $V_{td}$, and $s_{12}V_{cb}^*$, rescaled by a factor of $1/|s_{12}V_{cb}|$
so that the base 
is the unit length.
The coordinates of the vertices of the unitarity 
triangle are  $A(\rho,\eta)$, $B(1,0)$ and $C(0,0)$.}
corresponding to this measurement $\pm 1 \sigma$,
$\sin 2\phi_1=0.45^{+0.44}_{-0.45}$, 
together with  the constraints derived from  other measurements\cite{PDG}.
While the current statistical uncertainty does not allow  anything conclusive,
this preliminary result is consistent with the Standard Model prediction.
\begin{table}
\caption{List of systematic errors of $\sin 2\phi_1$}
\label{tab:systematics} 
\begin{tabular}{lll}
\hline
Source & $\sigma+$ & $\sigma-$\\
\hline
Wrong tag & $0.050$ & $-0.066$\\
$R(\Delta t)$  &$0.026$ &$ -0.025$ \\
Background shape&$0.029$& $-0.042$\\
Background fraction &$0.029$ & $-0.032$\\
$\tau_{B^0}$, $\Delta m_d$ & $0.005$ & $-0.006$\\
IP profile & $0.004$ & $-0.000$\\
\hline
Total & $+0.07$ & $-0.09$\\
\hline
\end{tabular}
\end{table}

\section*{Acknowledgments}

We are grateful to the conference organizers for their hospitality.
It is a pleasure to thank  the KEKB group
and the KEK computing research center.
We acknowledge support from the Ministry of Education, Science, Sports and
Culture of Japan and
the Japan Society for the Promotion of Science;
the Australian Research Council and the Australian Department of Industry,
Science and Resources;
the Department of Science and Technology of India;
the BK21 program of the Ministry of Education of Korea and
the Basic Science program of the Korea Science and Engineering Foundation;
the Polish State Committee for Scientific Research; 
the Ministry of Science and Technology of Russian Federation;
the National Science Council and the Ministry of Education of Taiwan;
the Japan-Taiwan Cooperative Program of the Interchange Association;
and  the U.S. Department of Energy.



\begin{thebibliography}{99}

\bibitem{Sanda}
H.~Quinn and A.I.~Sanda, Eur.Phys.J. C {\bf 15}, 625 (2000).

\bibitem{Belle}
Belle Collaboration, ``The Belle Detector," to be submitted to Nucl.Instr. and Methods.

\bibitem{Schrenk}
S.~Schrenk (Belle Collaboration), ``Studies of $B$ Meson Decays to Final States Containing 
Charmonium with Belle," in this Proceedings. 

\bibitem{Tajima}
H.~Tajima (Belle Collaboration), ``Measurement of Heavy Meson Lifetimes with Belle," {\it ibid}.
 
\bibitem{PDG}
Particle Data Group, D.E.~Groom {\it et al.}, Eur.Phys.J. C {\bf 15}, 1(2000).

\bibitem{paper285}
Belle Collaboration, ``Measurement of Polarization of $J/\psi$ in $B^0\to$ $J/\psi +K^{*0}$ and
$B^+\to$ $J/\psi+ K^{*+}$ Decays," Contributed paper (\#285) to
the XXXth International Conference on High Energy Physics, 
July 27 - August 2, 2000, Osaka, Japan. http://bsunsrv1.kek.jp/conferences 
/ichep2000.html

\end{thebibliography}
\end{document}